\documentclass[a4paper,11pt]{article}
\pdfoutput=1

\usepackage{jheppub}

\usepackage[abs]{overpic}

\usepackage{warpcol}

\usepackage[T1]{fontenc}

\RequirePackage{lineno}
\usepackage{epstopdf}
\newcommand{\too}{\rightarrow}
\newcommand{\EE}{e^+e^-}

\title{\boldmath Cross section measurements of the processes $\EE \too \omega\pi^{0}$ and $\omega\eta$ at center-of-mass energies between 3.773 and 4.701 GeV}

\collaborationImg{\includegraphics[width=.12\textwidth,origin=c,angle=90]{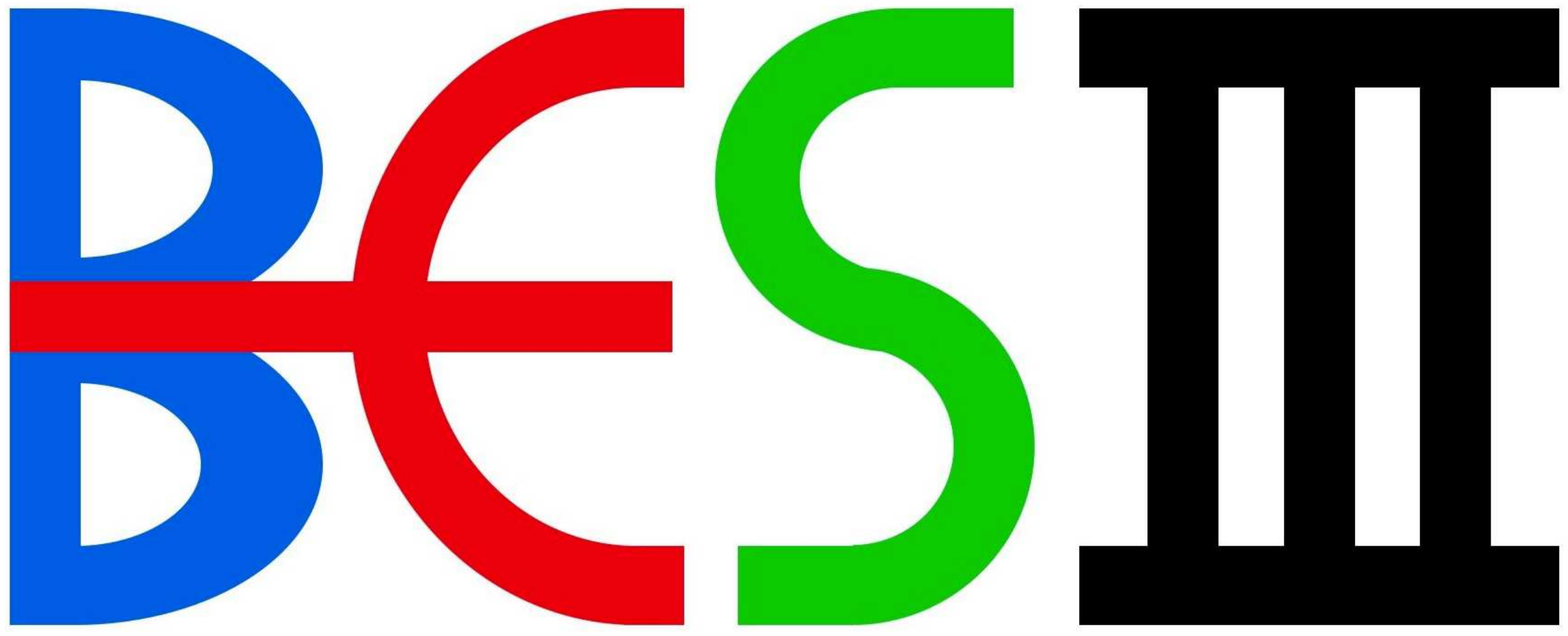}}

\collaboration{The BESIII collaboration}


\emailAdd{besiii-publications@ihep.ac.cn}

\abstract{

The Born cross sections of the processes $\EE\too\omega\pi^{0}$ and $\EE\too\omega\eta$ are measured at center-of-mass energies between 3.773 and 4.701~GeV using a total integrated luminosity of 22.7 fb$^{-1}$ collected with the BESIII detector operating at the BEPCII collider.
A simple $s^{-n}$ dependence for the continuum process can describe the measured Born cross sections.
No significant contributions from the $\psi(4160)$, $Y(4230)$, $Y(4360)$, $\psi(4415)$, $Y(4660)$ resonances are found, which indicates relative small branching fractions for these resonances into the $\omega\pi^{0}$ and $\omega\eta$ final states.}

\begin{document}
\maketitle
\flushbottom


\section{Introduction}
The recent discovery of several charmonium-like ($\emph{XYZ}$) states above the open-charm thresholds has attracted great experimental and theoretical interest~\cite{pdg}.  Due to the unexpected resonance parameters and decay patterns, they can not be described by conventional quark models, and are considered to be candidates for exotic states, such as hybrids, tetraquarks, and molecules~\cite{theory1, theory2, theory3}.

Since 2003, a series of charmonium-like states, such as $X$(3872)~\cite{x3872}, $Y(4260)$~\cite{y4260} and $Z_{c}$(3900)~\cite{zc1, zc2}, has been discovered. In particular, the vector charmonium-like state $Y(4260)$ was observed by the BaBar experiment in $e^{+}e^{-}\to\gamma_{ISR}\pi^{+}\pi^{-}J/\psi$ and confirmed by the CLEO and Belle experiments~\cite{intro4,intro5}. In addition, the $Y(4360)$ and $Y(4660)$ were also observed in $e^{+}e^{-}\to\gamma_{ISR}\pi^{+}\pi^{-}\psi(3686)$~\cite{y4360, y4660}.  Later, the BESIII experiment performed a dedicated scan for the channel $e^{+}e^{-}\to\pi^{+}\pi^{-}J/\psi$, revealing that the state previously identified as the $Y(4260)$ consists instead of two structures. The main component, with a mass of $M = (4222.0 \pm 3.1 \pm 1.4)$ MeV/$c^2$~\cite{intro6}, was renamed as the $Y(4230)$ by the Particle Data Group (PDG)~\cite{pdg}. The second structure is consistent with the previously observed $Y(4360)$. The $Y(4230)$ is also observed in the processes $\EE\too\omega\chi_{c0}$~\cite{intro7}, $\pi^{+}\pi^{-}h_{c}$~\cite{intro8}, $\pi^{+}\pi^{-}\psi(3686)$~\cite{intro9}, and $\pi^{+}D^{0}D^{*-}$~\cite{intro10}. Experimentally, the $Y$ states were mostly observed in channels with hidden or open charm states. Searches for the $Y$ states decaying into light hadron final states will improve our understanding of the $Y$ states. Several processes with light hadron final states have been measured by the BESIII experiment, such as $e^{+}e^{-}\to K_{S}^{0}K^{\pm}\pi^{\mp}\pi^{0}$ and $K_{S}^{0}K^{\pm}\pi^{\mp}\eta$~\cite{light1}, $p\bar{n}K^{0}_{S}K^{-} + c.c.$~\cite{light2}, $p\bar{p}\pi^{0}$~\cite{light3}, $p\bar{p}\eta$ and $p\bar{p}\omega$~\cite{light4}, but no significant charmonium-like structures are found. Further exploration of other light hadron final states is desirable to probe the nature of the charmonium-like states~\cite{hadcharm1, hadcharm2}. Recently, the processes of $e^{+}e^{-}\too\omega\pi^0$ and $\omega\eta$ have been reported at BESIII at $\sqrt{s}=2.00-3.08$ GeV~\cite{lowenergy}, in which two structures, $Y(2040)$ and $\phi(2170)$, are observed in respective line shape of cross section. The measurement of the cross sections for the two above processes at high energy is an extension to the same processes at lower energy region.

In this paper, we report measurements of the Born cross sections for the $e^{+}e^{-}\too\omega\pi^0$ and $\omega\eta$ processes at center-of-mass energies ($\sqrt{s}$) between 3.773 and 4.701~GeV with a total integrated luminosity of 22.7 fb$^{-1}$ and the subsequent search for $Y$ states or conventional charmonium states above the continuum contribution.

\section{BESIII detector and Monte Carlo simulation}
The BESIII detector is a magnetic spectrometer~\cite{besiii} located at the Beijing Electron
Positron Collider (BEPCII)~\cite{bepcii}. The
cylindrical core of the BESIII detector consists of a helium-based
multilayer drift chamber (MDC), a plastic scintillator time-of-flight
system (TOF), and a CsI (Tl) electromagnetic calorimeter (EMC),
which are all enclosed in a superconducting solenoidal magnet,
providing a 1.0~T magnetic field. The solenoid is supported by an
octagonal flux-return yoke with resistive plate chamber muon
identifier modules interleaved with steel. The acceptance of
charged particles and photons is 93\% over the $4\pi$ solid angle. The
charged-particle momentum resolution at $1~{\rm GeV}/c$ is
$0.5\%$, and the $\textrm{d}E/\textrm{d}x$ resolution is $6\%$ for the electrons
from Bhabha scattering. The EMC measures photon energies with a
resolution of $2.5\%$ ($5\%$) at $1$~GeV in the barrel (end cap)
region. The time resolution of the TOF barrel section is 68~ps, while that of the end cap section is 110~ps. The end cap TOF
system was upgraded in 2015 with multi-gap resistive plate chamber
technology, providing a time resolution of
60~ps~\cite{etof}; this improvement benefits 25 of the 34 energy points
used in this paper.

Simulated data samples produced with the {\sc geant4}-based~\cite{geant4} Monte Carlo (MC) package, which
includes the geometric description of the BESIII detector and the
detector response, are used to determine the detection efficiency
and to estimate the background contributions.
The simulation includes the beam
energy spread and initial-state radiation (ISR) in the $e^+e^-$
annihilations modeled with the generator {\sc kkmc}~\cite{KKMC}.
The ISR production of vector charmonium(-like) states and the continuum processes are incorporated also in {\sc kkmc}~\cite{KKMC}. The known decay modes are modeled with {\sc evtgen}~\cite{ref:evtgen}, using branching fractions summarized and averaged by the
PDG~\cite{pdg}, and the remaining unknown decays
from the charmonium states are generated with {\sc lundcharm}~\cite{ref:lundcharm}. Final state radiation from charged final state particles is incorporated with the {\sc photos} package~\cite{photos}.

Signal MC samples for $\EE \too \omega\pi^{0}$ and $\omega\eta$ are generated using  HELAMP (helicity amplitude model) and {\sc evtgen}~\cite{ref:evtgen} at each center-of-mass energy point.
The event selection criteria and the detection efficiencies are determined and studied based on signal MC samples of $1\times10^{5}$ signal events generated for each value of $\sqrt{s}$.
Detection efficiencies are determined by the ratio of the reconstructed event yields (after the selection criteria) to the number of the generated events.

\section{Event selection}
For each charged track, the distance of closest approach to the interaction point (IP) is required to be within $10$ cm in the beam direction and within 1 cm in the plane perpendicular to the beam direction. The polar angles ($\theta$) of the tracks must be within the fiducial volume of the MDC, $|\!\cos\theta|<0.93$. Photons are reconstructed from isolated showers in the EMC, which are at least $10^\circ$ away from the nearest charged track. The photon energy is required to be at least 25 MeV in the barrel region $(|\!\cos\theta|<0.80)$ or 50 MeV in the end cap region $(0.86<|\!\cos\theta|<0.92)$. To suppress electronic noise and energy depositions unrelated to the event, the EMC cluster timing from the reconstructed event start time is further required to satisfy $0\leq t \leq 700$ ns.

The final state $\omega$ is reconstructed via $\omega\too\pi^{+}\pi^{-}\pi^{0}$; this $\pi^0$ is referred to as the resonance $\pi^0$.
In addition there is a ``bachelor'' $\pi^0 \, (\eta)$.
All $\pi^0$ and $\eta$ are reconstructed via the decays to $\gamma \gamma$.  Given this topology, candidate events are required to have two charged tracks with zero net charge and at least four photons.
The flight time in the TOF and d$E$/d$x$ information in the MDC are combined to calculate particle identification (PID) likelihoods for the $\pi$, $K$, and $p$ hypotheses. For both charged tracks, it is required that the likelihood for a pion assignment is larger than that for both the kaon and proton hypotheses.

\begin{figure}[tbp]
\centering
\begin{overpic}[width=0.43\textwidth]{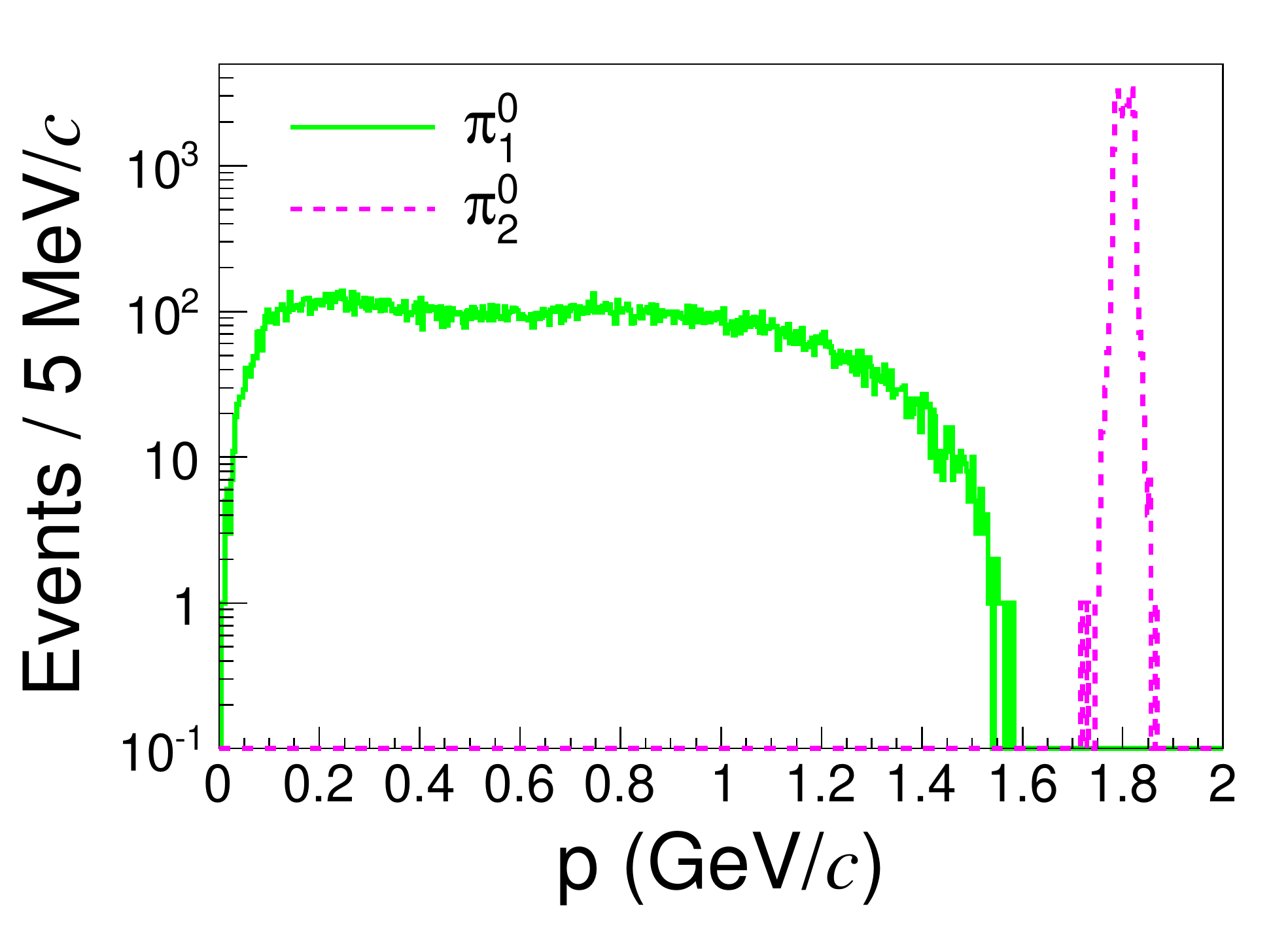}
\put(130,107){\LARGE{(a)}}
\end{overpic}
\begin{overpic}[width=0.43\textwidth]{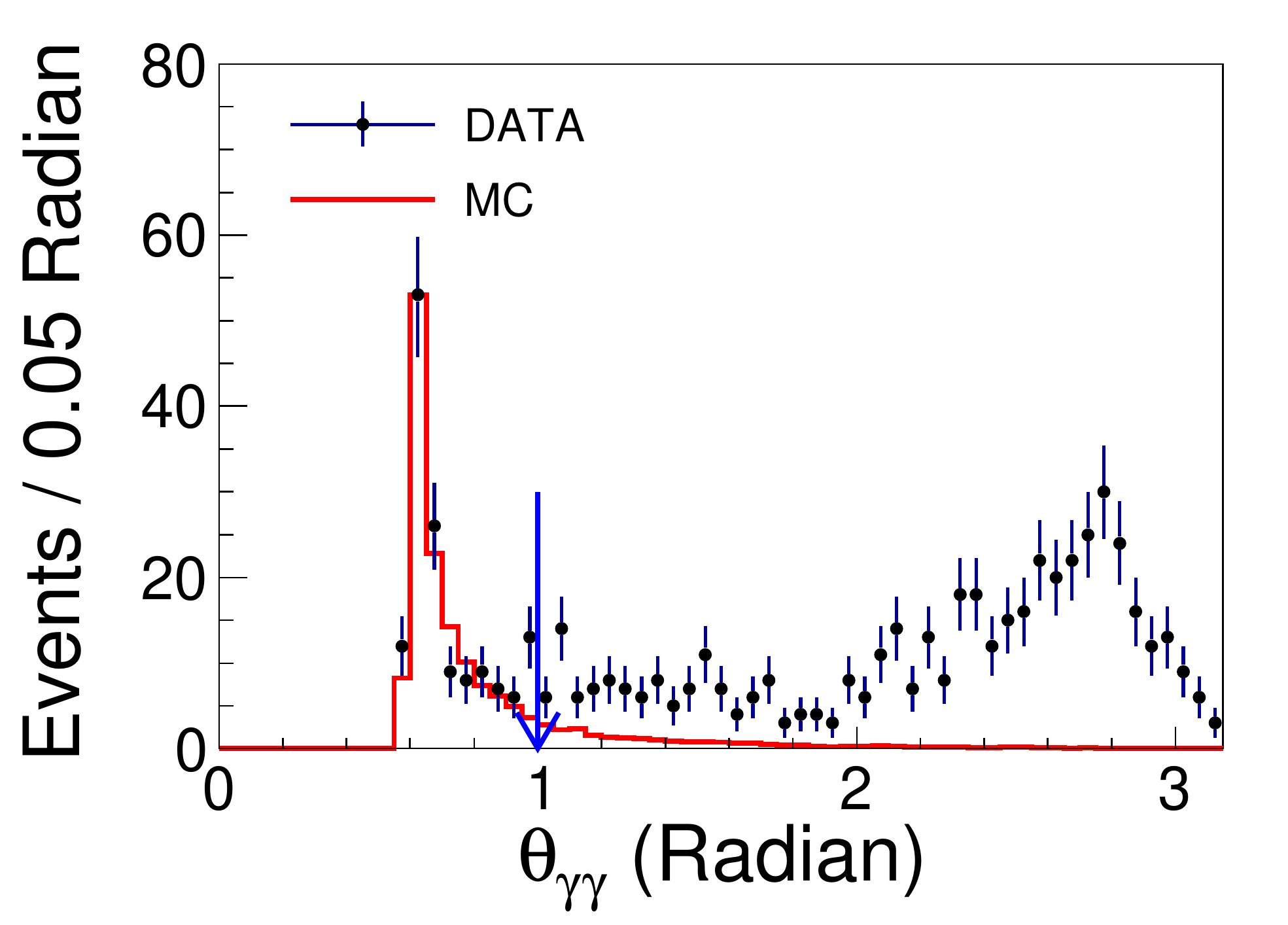}
\put(130,107){\LARGE{(b)}}
\end{overpic}
\caption{(a) Momentum distributions of the $\pi^{0}$ from signal MC in the lab-frame at $\sqrt{s}$ = 3.773 GeV for $\EE\too\omega\pi^{0}$. Here, $\pi^{0}_{1}$ is from $\omega$ resonance, and $\pi^{0}_{2}$ is the bachelor, directly from $\EE$. (b) The distribution of $\theta_{\gamma\gamma}$ for the $\eta\too\gamma\gamma$ at $\sqrt{s}$ = 3.773 GeV for $\EE\too\omega\eta$. Dots with error bars are data, the red solid line is signal MC, the vertical blue arrow indicates the requirement, $\theta_{\gamma\gamma} <1.0$ radian, that is used to select signal events.
}
\label{fig:pipi}
\end{figure}

A six-constraint (6C) kinematic fit is performed to the candidate events with $\EE\too\pi^{+}\pi^{-}\pi^0\pi^0, \pi^{+}\pi^{-}\pi^0\eta$ hypothesis. The total four-momentum is constrained to the initial four-momentum of the $e^+e^-$ system. The invariant mass of two photons from the $\omega$ resonance $\pi^{0}$ decay is constrained to the nominal $\pi^{0}$ mass~\cite{pdg} and the other two photons, from the bachelor $\pi^{0}$ ($\eta$) decay, are constrained to the nominal masses of the $\pi^{0}$ ($\eta$)~\cite{pdg}.
Multiple combinations arise from different photon pairings as well as events with more than four photon candidates.  The combination with the smallest $\chi^{2}_{\text{6C}}$ is chosen. For the two $\pi^0$ candidates in $\EE\too\omega\pi^{0}$, the momentum of the bachelor $\pi^{0}$ is larger than that of the $\pi^{0}$ from $\omega$ decay at all the energy points, allowing for separation of the  $\pi^0$ candidates. Figure~\ref{fig:pipi} (a) shows the momentum distributions of $\pi^{0}$ at $\sqrt{s}$ = 3.773 GeV for the signal MC.

After applying the above requirements, two more selection criteria are applied for background suppression. In order to study the non-resonant backgrounds, such as the background of $\EE\too\omega\gamma\gamma$, a five-constraint (5C) kinematic fit is performed on the selected candidate events.  This fit is simply the 6C kinematic fit with the mass constraint on the bachelor $\pi^0$ ($\eta$) removed. The $\chi^{2}_{\text{5C}}$ is required to be less than 60; further details are given in the next section. According to a study of inclusive MC samples with an event-type investigation tool, TopoAna~\cite{topoana}, the main background for $\EE\too\omega\eta$ is the ISR process $\EE\too\omega\gamma_{ISR}$.
To remove the background, the angle between the two photons from the $\eta$ in the laboratory frame, $\theta_{\gamma\gamma}$, is required to be less than 1 radian. Figure~\ref{fig:pipi} (b) shows the distribution of $\theta_{\gamma\gamma}$ at $\sqrt{s}$ = 3.773 GeV.

\section{Born cross section measurement}

\subsection{$\EE\too\omega\pi^{0}$}
\begin{figure}[htbp]
\begin{center}
\begin{overpic}[width=0.43\textwidth]{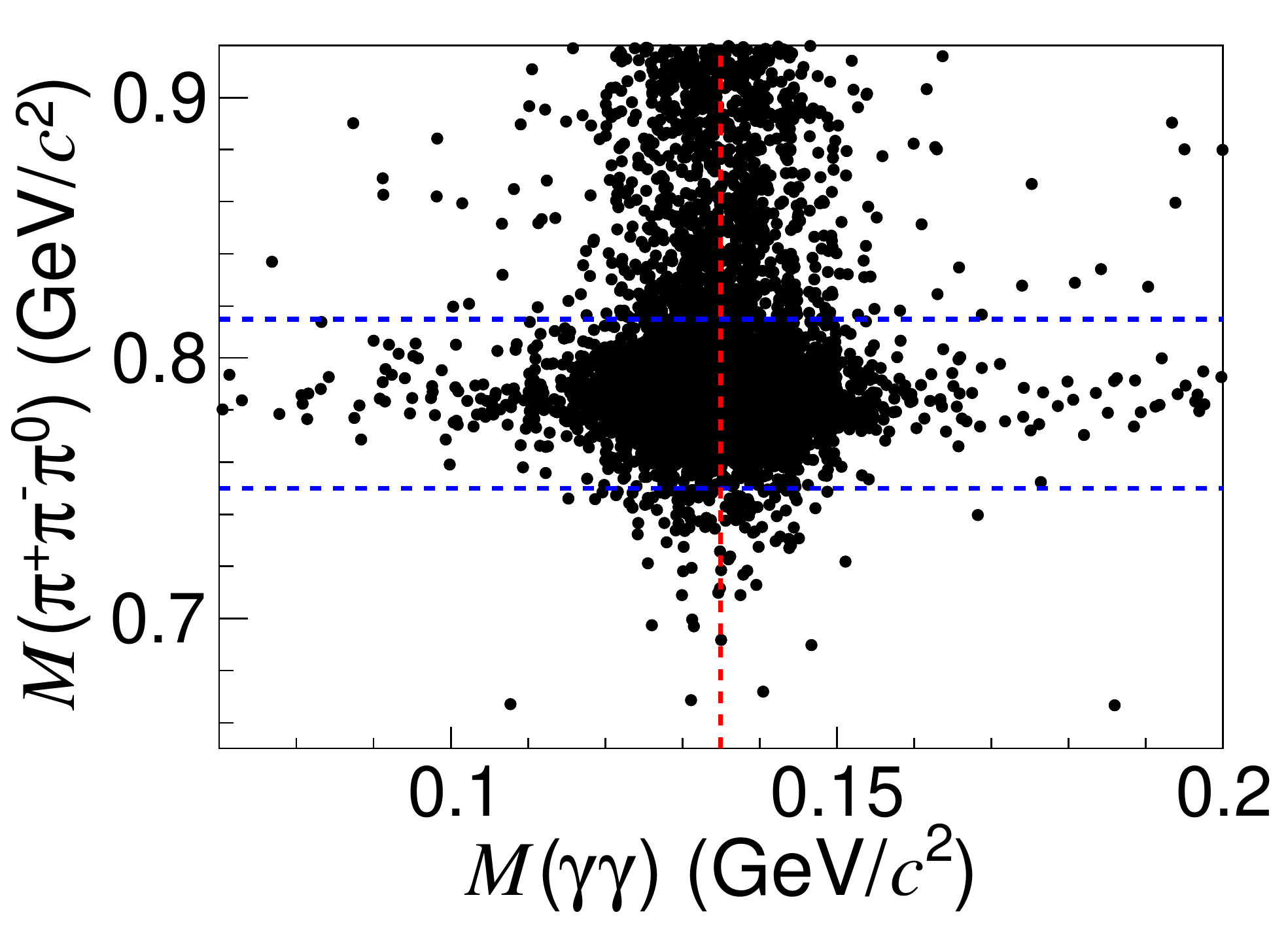}
\put(45,107){\LARGE{(a)}}
\end{overpic}
\begin{overpic}[width=0.43\textwidth]{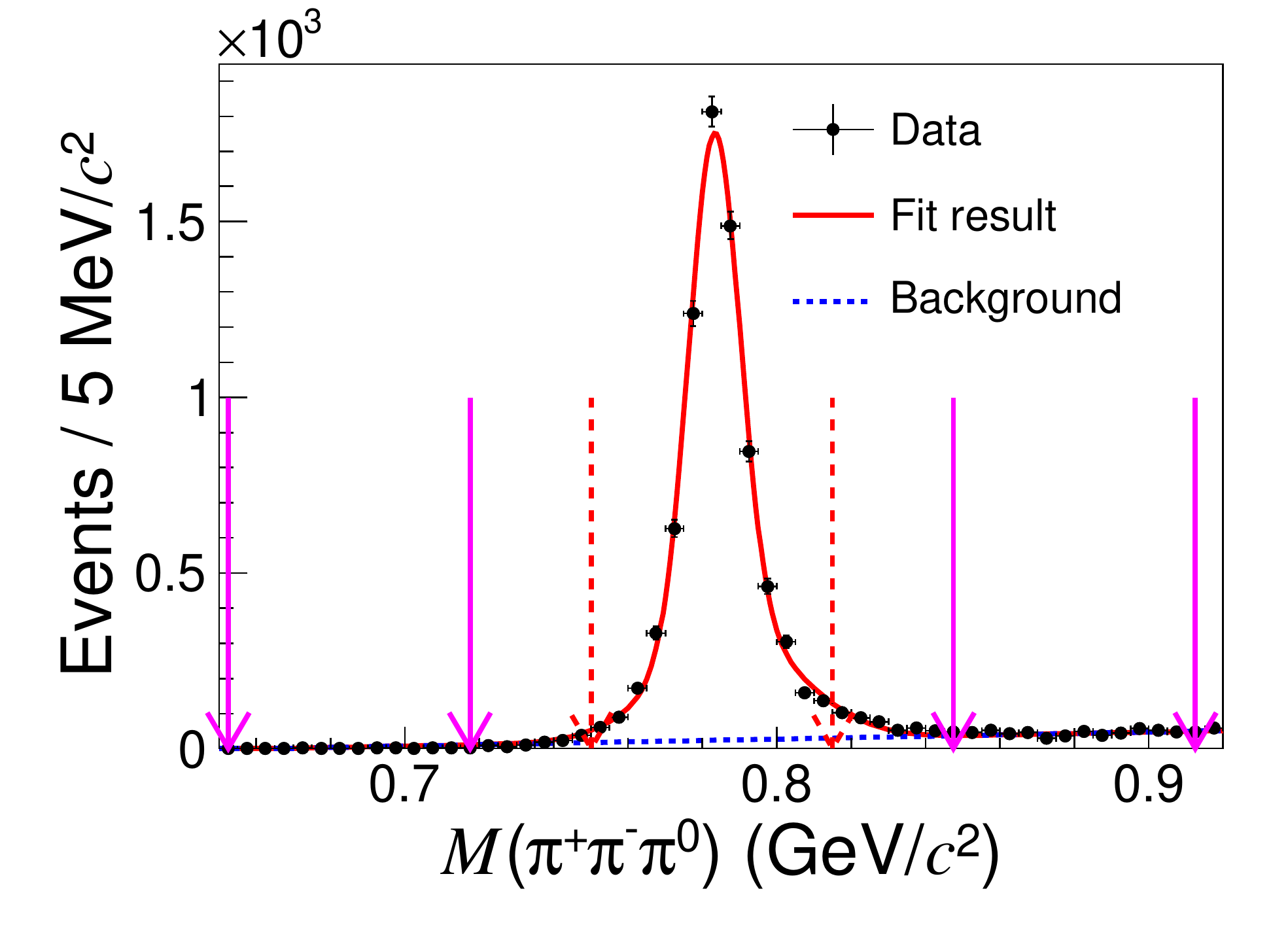}
\put(45,107){\LARGE{(b)}}
\end{overpic}
\caption{(a) $M(\pi^{+}\pi^{-}\pi^{0})$ versus $M(\gamma\gamma)$ for data at $\sqrt{s}$ = 3.773 GeV. The blue dashed lines mark the signal band of the $\omega$, and the red dashed line marks the mass of the $\pi^{0}$. (b) Fit to distribution of $M(\pi^{+}\pi^{-}\pi^{0})$ for the events in (a). The red dashed arrows mark the signal region of the $\omega$, and the pink solid arrows mark the sideband regions of the $\omega$.}
\label{fig:saomegapi0}
\end{center}
\end{figure}

Figure~\ref{fig:saomegapi0} (a) shows the distribution of $M(\pi^{+}\pi^{-}\pi^{0})$ versus $M(\gamma\gamma)$ after the 5C kinematic fit for data at $\sqrt{s}$ = 3.773 GeV. Here, $M(\gamma\gamma)$ is the invariant mass of the two photons from the bachelor $\pi^0$ decay. A clear $\omega\pi^0$ signal can be seen. Figure~\ref{fig:saomegapi0} (b) shows the distribution of $M(\pi^{+}\pi^{-}\pi^{0})$ for data at $\sqrt{s}$ = 3.773 GeV.
The $M(\pi^{+}\pi^{-}\pi^{0})$ spectrum is fit with a double-Gaussian function describing the signal and a linear function describing the background. Based on signal MC simulation, the $\omega$ signal region is defined as the mass range [0.7500, 0.8150]~GeV/$c^{2}$ in $M(\pi^{+}\pi^{-}\pi^{0})$, and is indicated by the horizontal dashed lines in Figure~\ref{fig:saomegapi0} (a). The sideband regions, defined as the range $[0.6525, 0.7175]\bigcup[0.8475, 0.9125]\,{\rm GeV/}c^{2}$ as indicated by the solid arrows in Figure~\ref{fig:saomegapi0} (b), are used to study the non-$\omega$ background.

\begin{figure}[htbp]
\begin{center}
\begin{overpic}[width=0.43\textwidth]{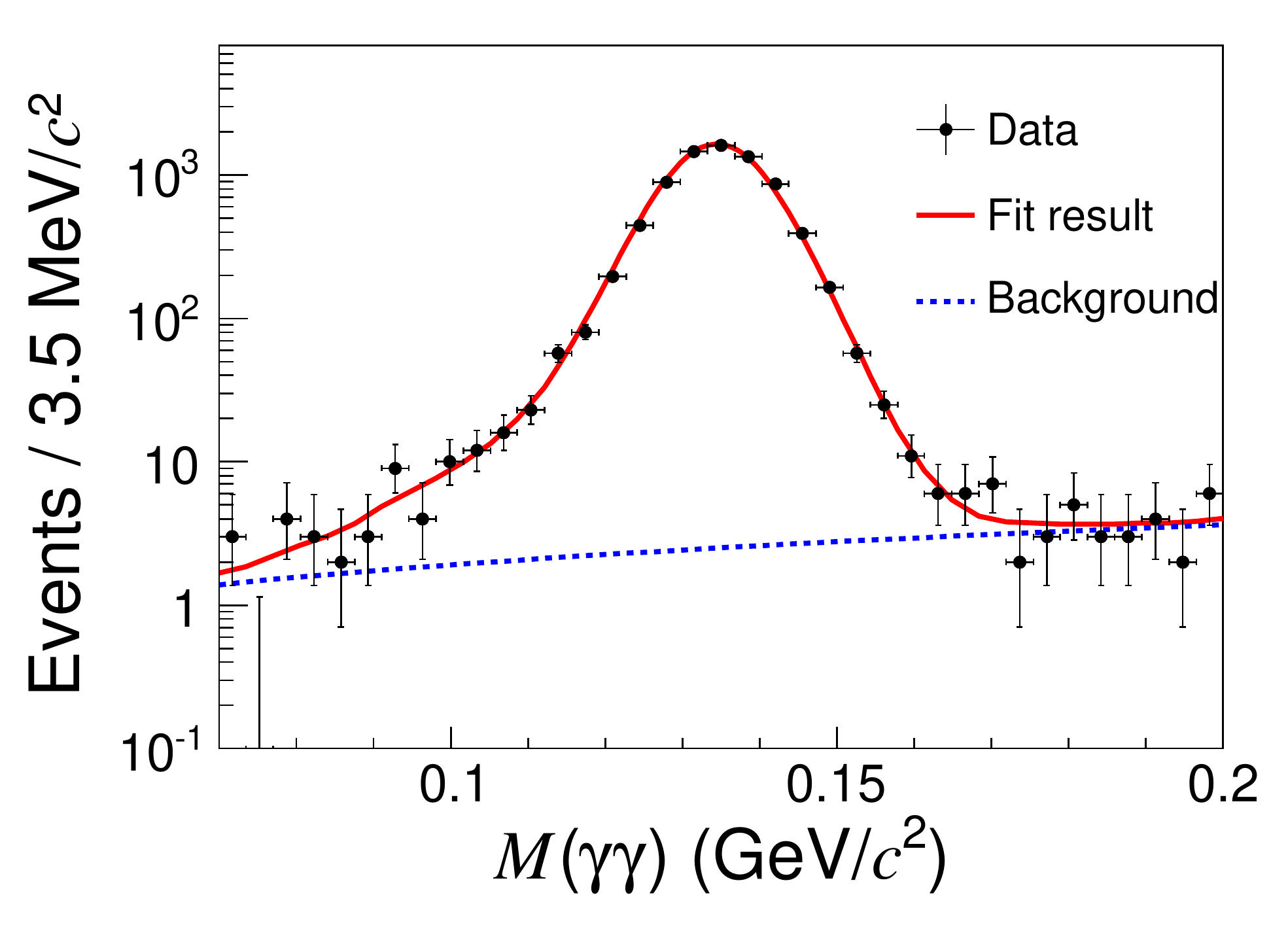}
\put(45,109){\LARGE{(a)}}
\end{overpic}
\begin{overpic}[width=0.43\textwidth]{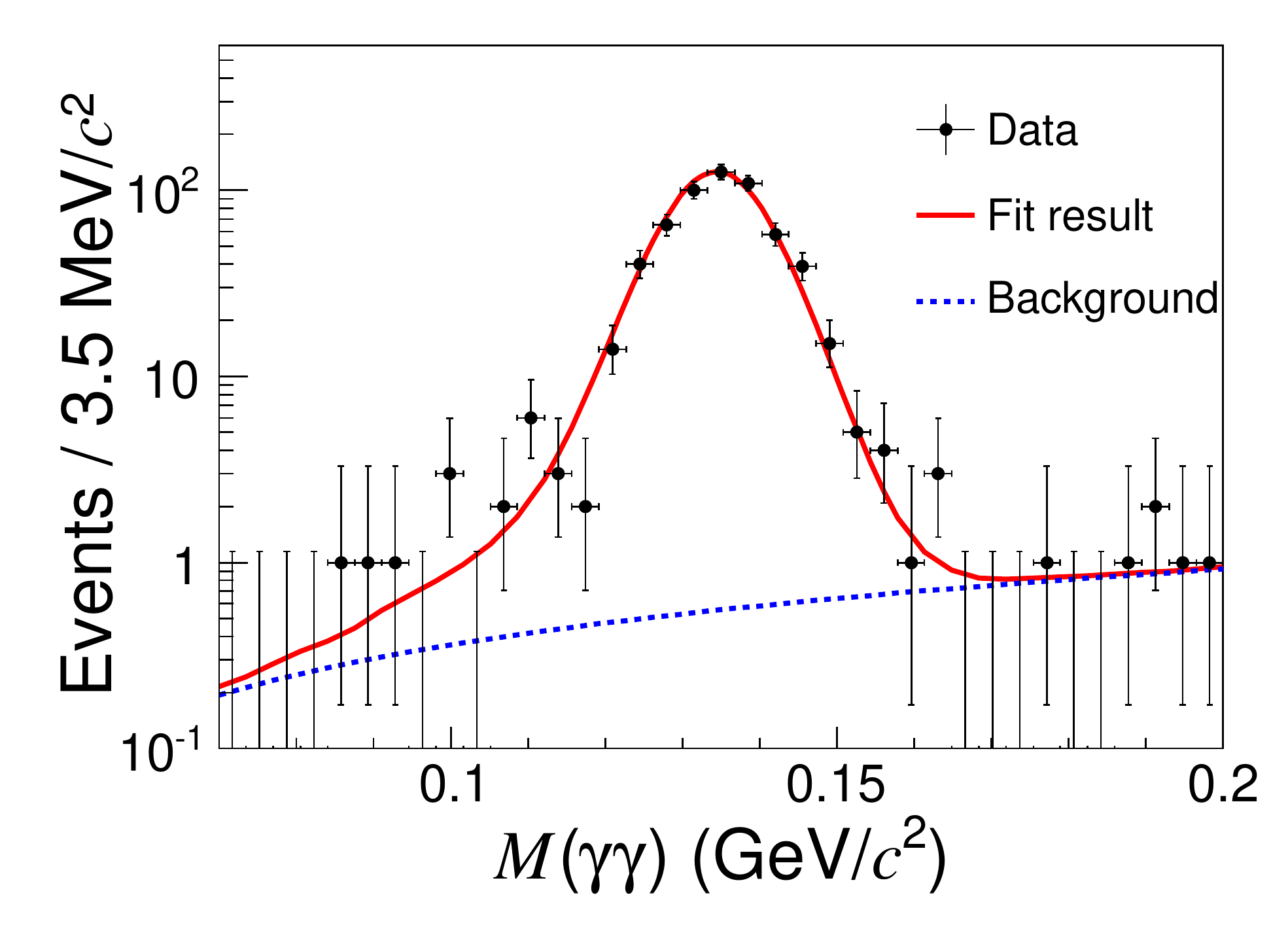}
\put(45,109){\LARGE{(b)}}
\end{overpic}
\caption{Fits to the distributions of $M(\gamma\gamma)$ for data from the $\omega$ signal region (a) and the $\omega$ sideband region (b) at $\sqrt{s}$ = 3.773 GeV.}
\label{fig:omegapi0}
\end{center}
\end{figure}

Figure~\ref{fig:omegapi0} shows the distributions of $M(\gamma\gamma)$ from the signal region (a) and the sideband region
(b) of the $\omega$ for data at $\sqrt{s}$ = 3.773 GeV. To account for non-$\omega$ backgrounds, the signal yields are obtained by unbinned maximum likelihood fits to the $\pi^0$ signal in the $M(\gamma\gamma)$ spectrum for events in the $\omega$ signal and sideband regions. The signal function comes from the MC-simulated shape, while the background shape is described by a linear background function. The fit results are shown in Figure~\ref{fig:omegapi0}. The net number of signal events is calculated by $N^\text{sig}=N^\text{obs}-N^\text{bkg}\cdot f_\text{scale}$, where $N^\text{obs}$ and $N^\text{bkg}$ are the fit result of $M(\gamma\gamma)$ in the $\omega$ signal and sideband regions, respectively, shown in Figure~\ref{fig:omegapi0}; $f_{scale}=0.5$ is the normalized factor between the $\omega$ signal and $\omega$ sideband region assuming the background shape is a smooth distribution.

\begin{figure}[htbp]
\begin{center}
\begin{overpic}[width=0.45\textwidth]{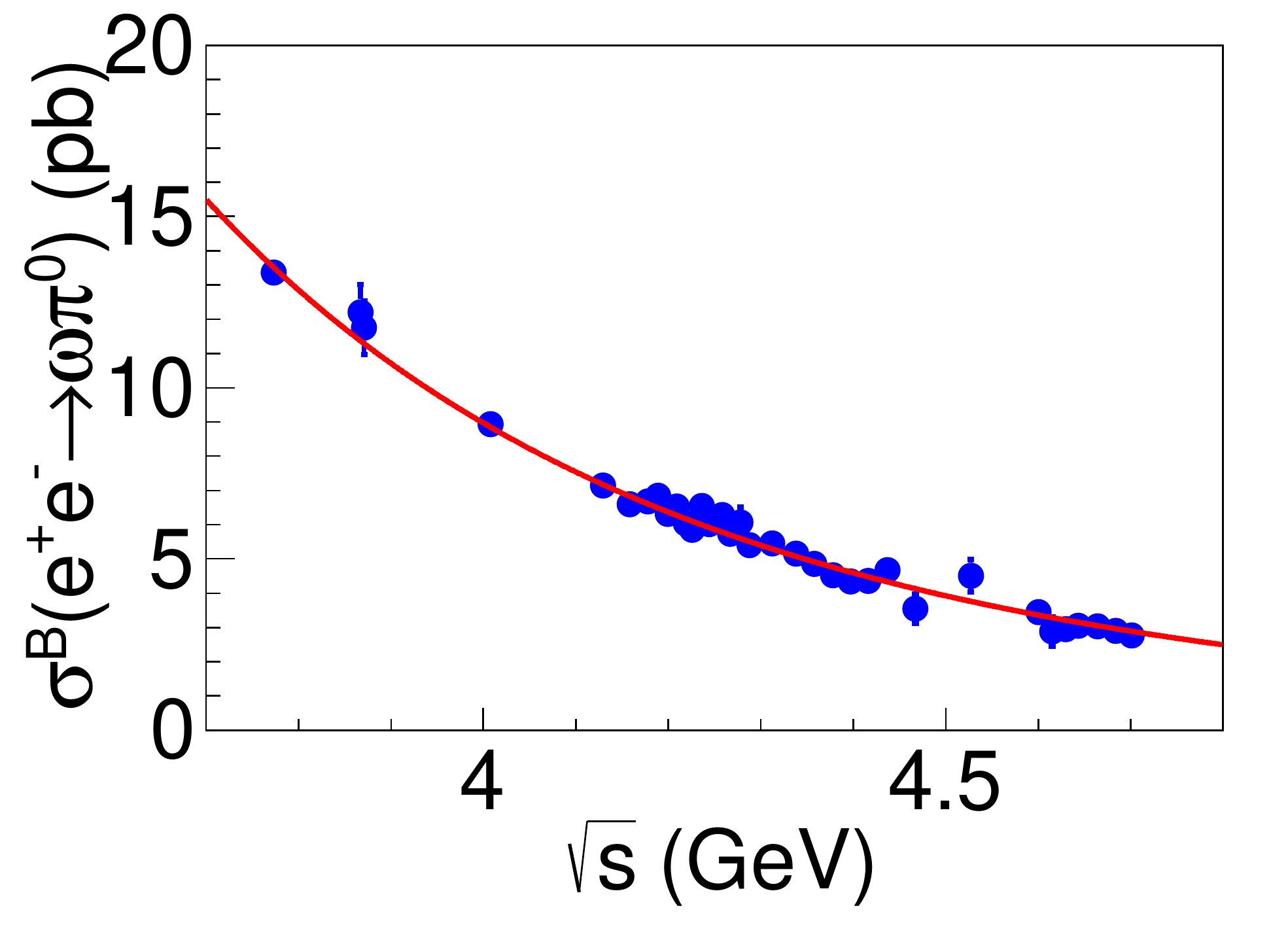}
\end{overpic}
\caption{Fit to the Born cross sections of $e^{+}e^{-}\rightarrow\omega\pi^{0}$ with function $a\cdot s^{-n}$.}
\label{fig:crosspi0}
\end{center}
\end{figure}

The Born cross section for $\EE\too\omega\pi^{0}$ is calculated using the following formula:
\begin{equation}
\begin{normalsize}
    \begin{aligned}
    \sigma^{\rm{Born}}\left(\sqrt{s}\right)=\frac{N^\text{sig}}{\mathcal{L}\cdot\epsilon\cdot\mathcal{B}\cdot\frac{1}{|1-\Pi|^{2}}\cdot(1+\delta(s))},
    \end{aligned}
    \label{con:sigma1}
\end{normalsize}
\end{equation}
where $N^\text{sig}$ is the number of signal events, $\mathcal{L}$ is the integrated luminosity, $\epsilon$ is the detection efficiency,
$\mathcal{B}$ is the product of the branching fractions in the full decay chain, $\mathcal{B}=\mathcal{B}(\omega\rightarrow \pi^{+}\pi^{-}\pi^{0})\cdot\mathcal{B}^{2}(\pi^{0}\rightarrow\gamma\gamma)\approx87.21\%$ taken from the PDG~\cite{pdg}, $\frac{1}{|1-\Pi|^{2}}$ is the vacuum polarization factor~\cite{vpfactor1, vpfactor2}, (1 + $\delta(s)$) is the ISR correction factor and is a function of $\sqrt{s}$. To obtain the ISR factor, we take the initial line shape of the observed cross section as an energy-dependent function $a\cdot s^{-n}$, which is used to describe the continuum process, and obtain the Born cross section iteratively until the results become stable within 1\% at all $\sqrt{s}$~\cite{ISRfactor}. The relevant numbers related to the Born cross section measurement are listed in Table~\ref{tab:omegapi0etaenergy}. Figure~\ref{fig:crosspi0} shows the energy-dependent Born cross section for $e^{+}e^{-}\rightarrow\omega\pi^{0}$, where $n=3.51\pm0.05$, and the goodness of fit is $\chi^{2}/n.d.f.=33.00/32$, where $n.d.f.$ is the number of degrees of freedom.

\begin{table*}[htbp]
  \centering
  \begin{scriptsize}
   \caption{The Born cross sections for $e^{+}e^{-}\rightarrow\omega\pi^{0}$ and $\omega\eta$,  together with integrated luminosities $\mathcal{L}_{\rm int}$, number of signal events $N^\text{sig}$, ISR factors 1+$\delta$(s), vacuum polarization factors $\frac{1}{|1-\Pi|^{2}}$, and efficiencies $\epsilon$. Here, $\sigma^{\rm{Born}}$ represents $\sigma^{\rm{Born}}_{\omega\pi^{0}}$ or $\sigma^{\rm{Born}}_{\omega\eta}$. For $N^\text{sig}$ and $\sigma^{\rm{Born}}$, errors are statistical only. The first values in brackets are for the process $e^{+}e^{-}\rightarrow\omega\pi^{0}$, and the second for the process $\EE\rightarrow\omega\eta$.}
  \label{tab:omegapi0etaenergy}
  \setlength{\tabcolsep}{0.5mm}{
  \begin{tabular}{c P{4.1} c c c c c}
  \hline
  \hline
  \multicolumn{1}{c}{$\sqrt{s}$(GeV)}  \ \  &   \multicolumn{1}{c}{$\mathcal{L}_{\rm int}$(pb$^{-1}$)}  &  \multicolumn{1}{c}{$N^\text{sig}$} & 1 + $\delta$(s) &  $\frac{1}{|1-\Pi|^{2}}$ &  $\epsilon(\%)$ &   $\sigma^{\rm{Born}}$(pb) \\
  \hline
  3.773 & 2931.8 & ($7335.2\pm88.9$, $96.8\pm10.6$)       & (1.0473, 1.0391)& 1.057 & (19.38, 15.01)& \ \ ($13.37\pm0.16$, $0.58\pm0.06$)\\
  3.867 & 108.9  & ($247.2\pm16.4$, $4.7_{-2.0}^{+2.6}$)  & (1.0678, 1.0570)& 1.051 & (19.02, 15.49)& ($12.21\pm0.81$, $0.72_{-0.31}^{+0.40}$)\\
  3.871 & 110.3  & ($240.4\pm16.0$, $4.6_{-2.0}^{+2.6}$)  & (1.0693, 1.0578)& 1.051 & (18.93, 15.58)& ($11.75\pm0.78$, $0.69_{-0.30}^{+0.39}$)\\
  4.008 & 482.0  & ($779.8\pm29.0$, $13.2_{-3.9}^{+4.5}$) & (1.0983, 1.0853)& 1.044 & (18.11, 15.43)& ($8.93\pm0.33$, $0.45_{-0.13}^{+0.15}$)\\
  4.129 & 401.5  & ($524.0\pm23.7$, $11.2_{-3.4}^{+4.0}$) & (1.1248, 1.1088)& 1.053 & (17.69, 14.88)& ($7.14\pm0.32$, $0.46_{-0.13}^{+0.15}$)\\
  4.158 & 408.7  & ($491.6\pm23.2$, $5.0_{-2.6}^{+3.2})$  & (1.1311, 1.1148)& 1.054 & (17.56, 15.10)& ($6.59\pm0.31$, $0.20_{-0.10}^{+0.13}$)\\
  4.178 & 3194.5 & ($3840.7\pm63.8$, $59.5\pm9.3$)        & (1.1352, 1.1190)& 1.055 & (17.24, 14.97)& \ \ \  ($6.68\pm0.11$, $0.30\pm0.05$)\\
  4.189 & 526.7  & ($649.0\pm26.6$, $15.8_{-3.9}^{+4.6}$) & (1.1386, 1.1207)& 1.056 & (17.21, 14.84)& ($6.84\pm0.28$, $0.49_{-0.12}^{+0.14}$)\\
  4.199 & 526.0  & ($601.1\pm25.5$, $13.8_{-3.5}^{+4.1}$) & (1.1398, 1.1224)& 1.057 & (17.24, 14.93)& ($6.31\pm0.27$, $0.43_{-0.11}^{+0.13}$)\\
  4.209 & 517.1  & ($603.9\pm25.5$, $19.1_{-4.2}^{+4.9}$) & (1.1421, 1.1242)& 1.057 & (16.99, 14.64)& ($6.53\pm0.28$, $0.61_{-0.13}^{+0.16}$)\\
  4.219 & 514.6  & ($555.3\pm24.7$, $6.7_{-2.4}^{+3.1}$)  & (1.1439, 1.1257)& 1.057 & (16.94, 14.83)& ($6.04\pm0.27$, $0.21_{-0.08}^{+0.10}$)\\
  4.226 & 1056.4 & ($1134.3\pm34.7$, $23.4_{-4.9}^{+5.6}$)& (1.1454, 1.1278)& 1.057 & (17.38, 15.05)& ($5.85\pm0.18$, $0.35_{-0.07}^{+0.08}$)\\
  4.236 & 530.3  & ($624.3\pm26.1$, $3.7_{-1.9}^{+2.6}$)  & (1.1488, 1.1296)& 1.056 & (16.99, 14.77)& ($6.55\pm0.27$, $0.11_{-0.06}^{+0.08}$)\\
  4.244 & 538.1  & ($580.3\pm25.2$, $11.0_{-3.3}^{+4.0}$) & (1.1504, 1.1320)& 1.056 & (16.94, 14.65)& ($6.01\pm0.26$, $0.34_{-0.10}^{+0.12}$)\\
  4.258 & 828.4  & ($914.5\pm31.6$, $19.2_{-4.5}^{+5.1}$) & (1.1535, 1.1339)& 1.054 & (16.59, 14.52)& ($6.28\pm0.22$, $0.38_{-0.09}^{+0.10}$)\\
  4.267 & 531.1  & ($537.2\pm24.1$, $7.0_{-2.6}^{+3.2}$)  & (1.1561, 1.1359)& 1.053 & (16.64, 15.06)& ($5.73\pm0.26$, $0.21_{-0.08}^{+0.10}$)\\
  4.278 & 175.7  & ($190.4\pm14.3$, $3.6_{-1.8}^{+2.5}$)  & (1.1582, 1.1375)& 1.053 & (16.82, 14.50)& ($6.06\pm0.45$, $0.34_{-0.17}^{+0.24}$)\\
  4.288 & 502.4  & ($486.1\pm22.8$, $10.7_{-3.3}^{+3.9}$) & (1.1609, 1.1396)& 1.053 & (16.77, 14.68)& ($5.41\pm0.25$, $0.35_{-0.11}^{+0.13}$)\\
  4.312 & 501.2  & ($494.8\pm23.2$, $15.8_{-3.9}^{+4.5}$) & (1.1662, 1.1453)& 1.052 & (16.89, 14.69)& ($5.46\pm0.26$, $0.51_{-0.13}^{+0.15}$)\\
  4.338 & 505.0  & ($463.9\pm22.4$, $5.1_{-2.4}^{+3.0}$)  & (1.1722, 1.1513)& 1.051 & (16.58, 14.58)& ($5.16\pm0.25$, $0.16_{-0.08}^{+0.10}$)\\
  4.358 & 543.9  & ($477.1\pm22.6$, $9.0_{-3.2}^{+3.8}$)  & (1.1770, 1.1558)& 1.051 & (16.77, 14.62)& ($4.85\pm0.23$, $0.27_{-0.10}^{+0.11}$)\\
  4.378 & 522.7  & ($420.6\pm21.5$, $8.4_{-3.2}^{+3.8}$)  & (1.1819, 1.1591)& 1.052 & (16.43, 14.49)& ($4.52\pm0.23$, $0.26_{-0.10}^{+0.12}$)\\
  4.397 & 507.8  & ($391.7\pm20.5$, $10.0_{-3.2}^{+3.9}$) & (1.1858, 1.1626)& 1.051 & (16.34, 14.45)& ($4.34\pm0.23$, $0.32_{-0.11}^{+0.13}$)\\
  4.416 & 1043.9 & ($814.7\pm29.7$, $13.1_{-3.8}^{+4.4}$) & (1.1903, 1.1670)& 1.053 & (16.38, 14.38)& ($4.36\pm0.16$, $0.20_{-0.06}^{+0.07}$)\\
  4.437 & 569.9  & ($471.9\pm22.7$, $11.9_{-3.4}^{+4.1}$) & (1.1950, 1.1709)& 1.054 & (16.15, 14.29)& ($4.67\pm0.22$, $0.34_{-0.10}^{+0.12}$)\\
  4.467 & 111.1  & ($70.2\pm8.6$, $0.3^{+1.6}_{-0.3}$)    & (1.2029, 1.1773)& 1.055 & (16.08, 14.27)& ($3.55\pm0.43$, $0.04^{+0.23}_{-0.04}$)\\
  4.527 & 112.1  & ($90.0\pm9.6$, $0.0^{+1.1}_{-0.0}$)    & (1.2177, 1.1886)& 1.055 & (15.87, 14.16)& ($4.51\pm0.48$, $0.00^{+0.16}_{-0.00}$)\\
  4.600 & 586.9  & ($347.3\pm19.5$, $7.5_{-2.8}^{+3.5}$)  & (1.2354, 1.2055)& 1.055 & (15.13, 13.70)& ($3.44\pm0.19$, $0.21_{-0.08}^{+0.10}$)\\
  4.615 & 102.5  & ($51.2\pm7.6$, $0.0^{+1.2}_{-0.0}$)    & (1.2386, 1.2087)& 1.055 & (15.20, 13.63)& ($2.88\pm0.43$, $0.00^{+0.19}_{-0.00}$)\\
  4.630 & 511.1  & ($259.5\pm17.1$, $3.0_{-1.6}^{+2.3}$)  & (1.2434, 1.2119)& 1.055 & (15.04, 13.61)& ($2.95\pm0.19$, $0.10_{-0.05}^{+0.07}$)\\
  4.643 & 541.4  & ($285.4\pm17.8$, $1.0^{+1.7}_{-1.0}$)  & (1.2466, 1.2158)& 1.055 & (15.09, 13.34)& ($3.05\pm0.19$, $0.03^{+0.05}_{-0.03}$)\\
  4.664 & 523.6  & ($273.3\pm17.1$, $4.5_{-2.0}^{+2.7}$)  & (1.2522, 1.2191)& 1.055 & (14.94, 13.75)& ($3.03\pm0.19$, $0.14_{-0.06}^{+0.08}$)\\
  4.684 & 1631.7 & ($814.9\pm29.8$, $17.5_{-4.5}^{+5.2}$) & (1.2569, 1.2226)& 1.055 & (14.89, 13.53)& ($2.90\pm0.11$, $0.18_{-0.05}^{+0.05}$)\\
  4.701 & 526.2  & ($247.4\pm16.5$, $7.7_{-2.7}^{+3.3}$)  & (1.2618, 1.2271)& 1.055 & (14.65, 13.55)& ($2.76\pm0.18$, $0.24_{-0.08}^{+0.10}$)\\
  \hline
  \hline
  \end{tabular}}
  \end{scriptsize}
\end{table*}

The potential contribution from conventional charmonium or charmonium-like states, $\psi(4160)$, $Y(4230)$, $Y(4360)$, $\psi(4415)$, and $Y(4660)$, is investigated by fitting the dressed cross section using the coherent sum of the continuum and an additional charmonium(-like) state amplitude. The corresponding fit function~\cite{photon} is expressed as
\begin{equation}
\begin{normalsize}
\begin{aligned}
   &\sigma^{\rm{D}}\left(\sqrt{s}\right) = \left|\left(a\, \cdot s^{-n}\right)^{1/2}+\frac{\sqrt{12\pi\Gamma_{ee}\mathcal{B}(\omega\pi^{0})\Gamma}}{s-M^{2}+iM\Gamma}\left({\frac{PS(\sqrt{s})}{PS(M)}}\right)^{3/2}e^{i\phi} \right|^{2},
\end{aligned}
 \label{con:sigma2}
 \end{normalsize}
\end{equation}
where $\sigma^{\rm{D}}(\sqrt{s})=\sigma^{{\rm{Born}}}(\sqrt{s})/|1-\Pi|^{2}$, is the dressed cross section, $\phi$ is the phase angle between the amplitude of the continuum process and the charmonium(-like) state, $PS(\sqrt{s})$ is the two-body phase space factor, $\Gamma_{ee}$ is the $\EE$ partial width, $\mathcal{B}(\omega\pi^0)$ is the branching fraction of charmonium(-like) decays to $\omega\pi^{0}$ final state,  $M$ and $\Gamma$ are the mass and width of charmonium(-like) state, which are fixed to their nominal values~\cite{pdg}. In this fit, $\Gamma_{ee}\mathcal{B}(\omega\pi^0)$, $\phi$, $n$, $a$ are the free parameters.

To examine the significance of the potential charmonium(-like) state, the fit is repeated using only the continuum amplitude. If using a continuum amplitude to fit $\sigma^{\rm{D}}$, the goodness of fit is $\chi^{2}/n.d.f.=32.65/32$. The significance is calculated taking into account the difference in likelihood value and the change in the number of degrees of freedom from the two fits, and the systematic uncertainty is not considered in the significance calculation. Table~\ref{tab:omegaetapi0cross} lists the fit parameters, the statistical significance and $\chi^{2}/n.d.f.$ for additional charmonium(-like) states. No obvious structure is found in the process of $e^{+}e^{-}\rightarrow\omega\pi^{0}$.

\begin{table*}[htbp]
  \centering
  \caption{Results of the fits to the dressed cross section $\sigma^\text{D}(\sqrt{s})$.
  ``Solution I'' represents the constructive solution, and ``Solution II'' represents the destructive solution. The uncertainty is statistical only.}
  \label{tab:omegaetapi0cross}
   \centering
   \begin{scriptsize}
   \setlength{\tabcolsep}{0.7mm}{
    \begin{tabular}{c c c cccccc}
  \hline
  \hline
   \ \ Channel \ \ &  \ \ Resonance \ \ & \ \  \  \  $\Gamma_{ee}\cdot\mathcal{B}$ (eV)  Solution I \  \  \ \ & \ \   \   \  \  $\Gamma_{ee}\cdot\mathcal{B}$ (eV) Solution II \ \ \  \  & \ \ Significance \ \ & \ \ \ $\chi^{2}/n.d.f.$ \ \ \\
  \hline
 & $\psi(4160)$ & $(2.31\pm2.92)\times10^{-4}$ & $2.29\pm0.02$ & 1.2$\sigma$ & $29.78/30$\\
 & $Y(4230)$ & $(2.96\pm7.61)\times10^{-5}$ & $1.39\pm0.01$ & 0.3$\sigma$ & $32.06/30$ \\
   $\omega\pi^{0}$
 & $Y(4360)$ & $(5.26\pm5.18)\times10^{-4}$ & $2.51\pm0.03$ & 1.5$\sigma$ & $28.47/30$ \\
 & $\psi(4415)$ & $(5.13\pm15.21)\times10^{-5}$& $1.54\pm0.02$ & 0.3$\sigma$ & $32.18/30$\\
 & $Y(4660)$ & $(3.97\pm4.82)\times10^{-4}$ & $1.23\pm0.02$ & 1.2$\sigma$ & $29.77/30$ \\
  \hline
 & $\psi(4160)$ & $(4.33\pm 7.75)\times10^{-4}$ & $(1.14\pm0.07)\times10^{-1}$ &   0.8$\sigma$ & $38.73/30$ \\
 & $Y(4230)$ & $(2.41\pm 4.21)\times10^{-4}$ & $(6.99\pm0.52)\times10^{-2}$ & 0.6$\sigma$ & $39.09/30$ \\
   $\omega\eta$
 & $Y(4360)$ & $(7.82\pm13.36)\times10^{-4}$ & $(1.33\pm0.14)\times10^{-1}$ & 0.7$\sigma$ & $38.98/30$ \\
 & $\psi(4415)$ & $(9.66\pm16.16)\times10^{-4}$ & $(7.74\pm1.10)\times10^{-2}$ &  0.8$\sigma$ & $38.73/30$ \\
 & $Y(4660)$ & $(3.45\pm 3.22)\times10^{-3}$ & $(5.02\pm0.97)\times10^{-2}$ &   2.3$\sigma$ & $32.71/30$ \\
  \hline
  \hline
   \end{tabular}}
   \end{scriptsize}
\end{table*}

\subsection{$\EE\too\omega\eta$}
\begin{figure}[htbp]
\begin{center}
\begin{overpic}[width=0.43\textwidth]{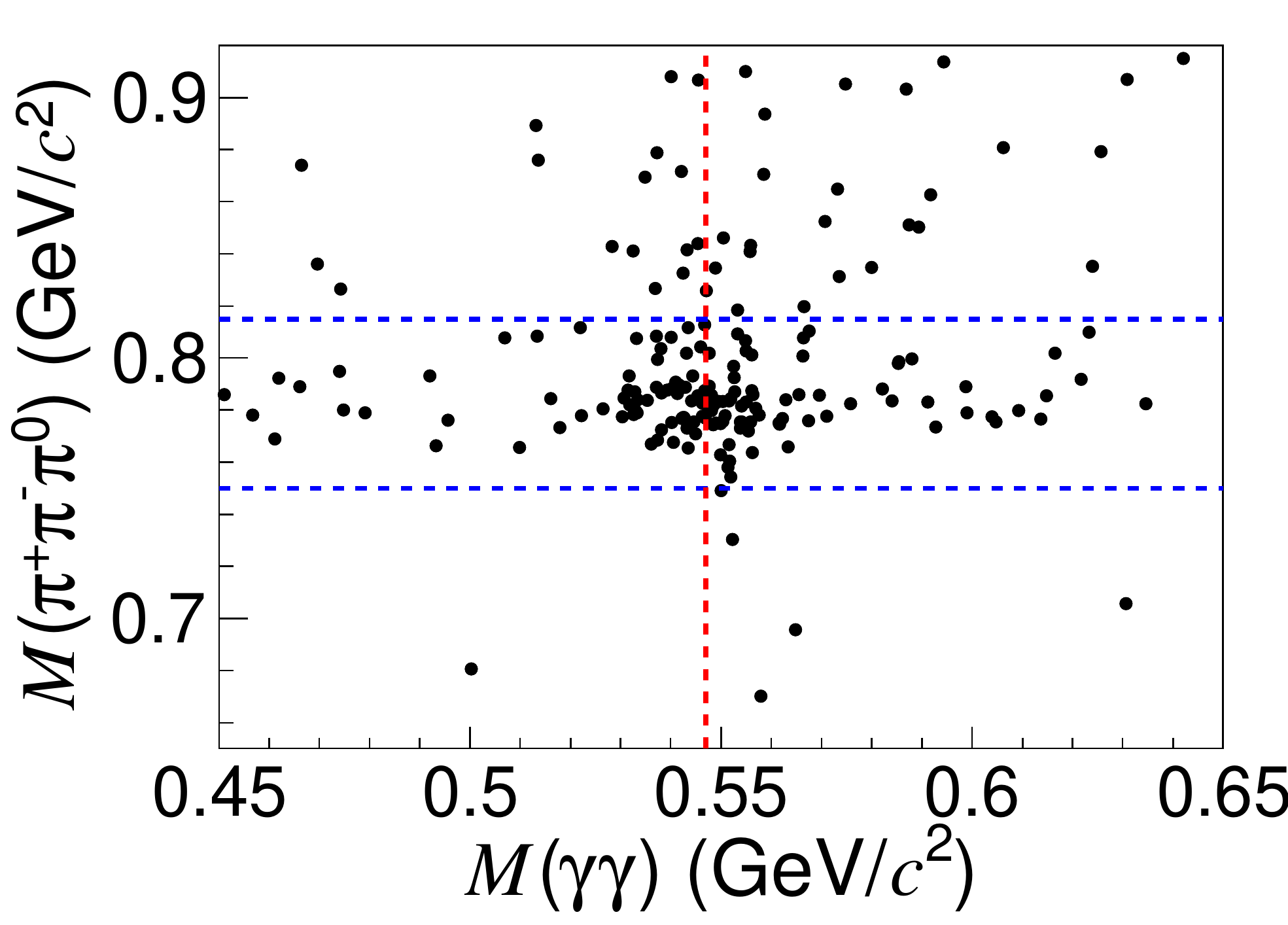}
\put(45,107){\LARGE{(a)}}
\end{overpic}
\begin{overpic}[width=0.43\textwidth]{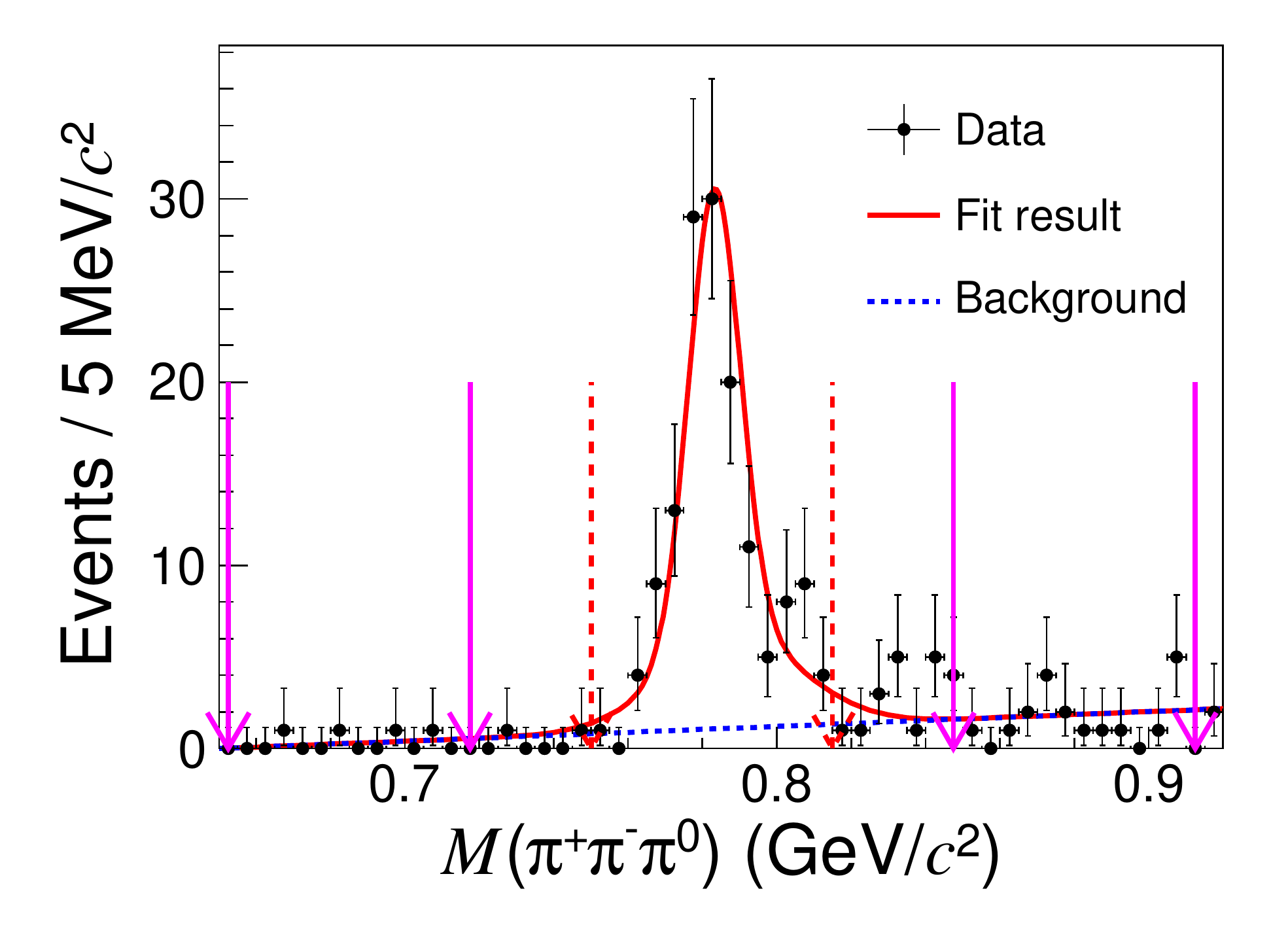}
\put(45,107){\LARGE{(b)}}
\end{overpic}
\caption{(a) $M(\pi^{+}\pi^{-}\pi^{0})$ versus $M(\gamma\gamma)$ for data at $\sqrt{s}$ = 3.773 GeV for $\EE\too\omega\eta$. The blue dashed lines mark the signal bands of $\omega$, and the red dashed line mark the known mass of $\eta$. (b) Fit to distribution of $M(\pi^{+}\pi^{-}\pi^{0})$ for the events in (a). The red dashed arrows mark the signal region of the $\omega$, and the pink solid arrows mark the sideband regions of the $\omega$.}
\label{fig:omegaeta1}
\end{center}
\end{figure}

Figure~\ref{fig:omegaeta1} (a) shows the distribution of $M(\pi^{+}\pi^{-}\pi^{0})$ versus $M(\gamma\gamma)$ (from $\eta$) for data at $\sqrt{s}$ = 3.773 GeV. Clear $\omega\eta$ signals are observed. Figure~\ref{fig:omegaeta1} (b) shows the distribution of $M(\pi^{+}\pi^{-}\pi^{0})$ for data at $\sqrt{s}$ = 3.773 GeV, the fitting method is same as Figure~\ref{fig:saomegapi0} (b), but the parameters of the signal shape are fixed to those from Figure~\ref{fig:saomegapi0} (b) due to the low statistics. The $\omega$ signal region is defined as the mass range $[0.7500,0.8150]\,{\rm GeV}/c^{2}$ in $M(\pi^{+}\pi^{-}\pi^{0})$ and is indicated by the horizontal dashed lines. The sideband regions, defined as the range $[0.6525,0.7175]\bigcup[0.8475,0.9125]\,{\rm GeV}/c^{2}$, are used to study the non-$\omega$ background.

\begin{figure}[htbp]
\begin{center}
\begin{overpic}[width=0.45\textwidth]{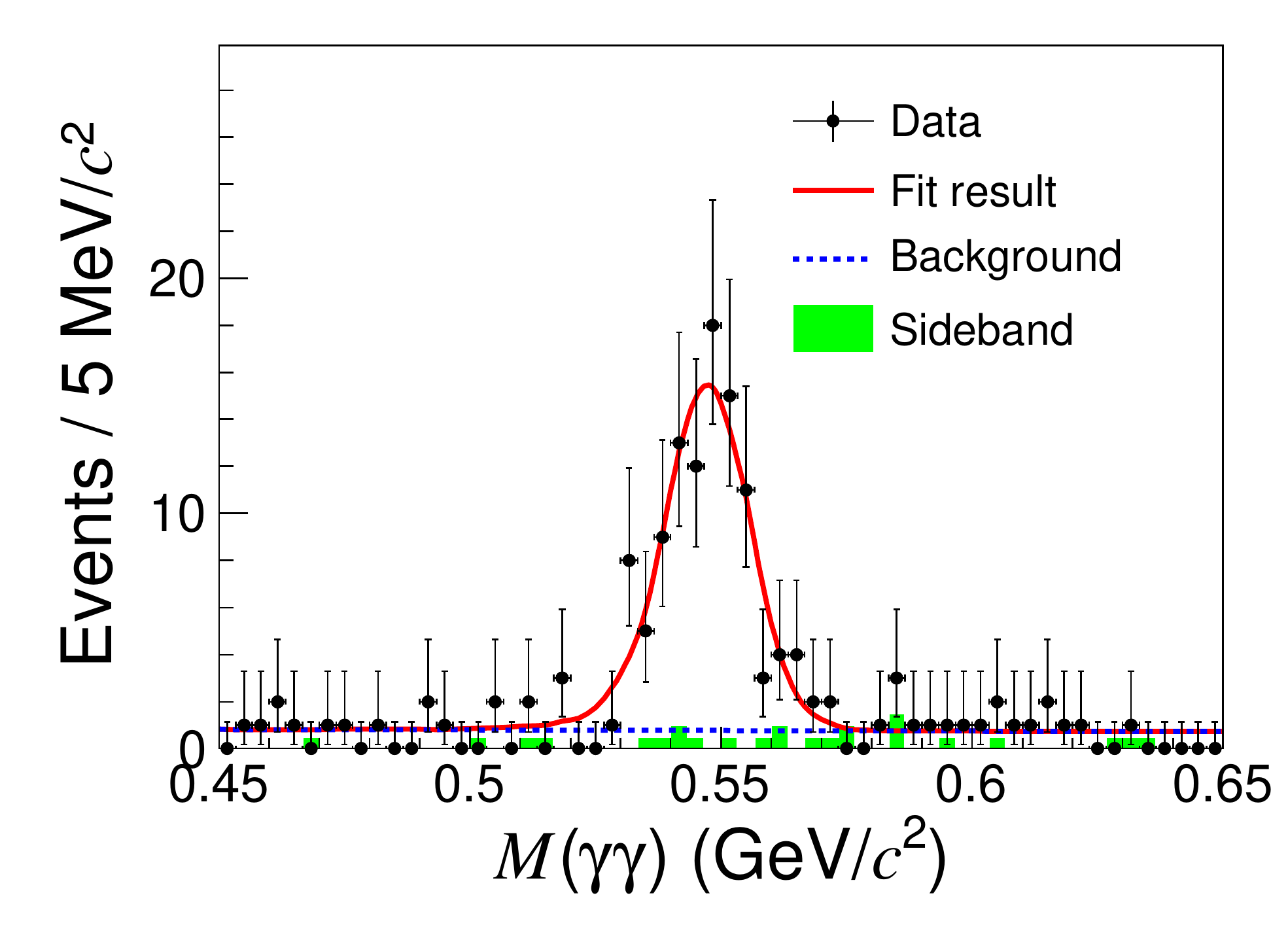}
\end{overpic}
\caption{The $M(\gamma\gamma)$ invariant mass distribution for the data at $\sqrt{s}$ = 3.773 GeV for $\EE\too\omega\eta$. The red solid line is the fit to the data and the blue dashed line is the background component. The green shaded histogram corresponds to the normalized background events from
the $\omega$ sideband region. }
\label{fig:omegaeta2}
\end{center}
\end{figure}

Figure~\ref{fig:omegaeta2} shows the distributions of $M(\gamma\gamma)$ for data at $\sqrt{s}=$3.773 GeV. No obvious non-$\omega$ events are seen as indicated by the very small sideband contribution in the green shaded histogram. The yield of signal events is obtained by an unbinned maximum likelihood fit to the $\eta$ signal in the $M(\gamma\gamma)$ spectrum for events in the $\omega$ signal region. The signal function is described by the MC-simulated shape, and the background shape is described by a linear function.

\begin{figure}[htbp]
\begin{center}
\begin{overpic}[width=0.45\textwidth]{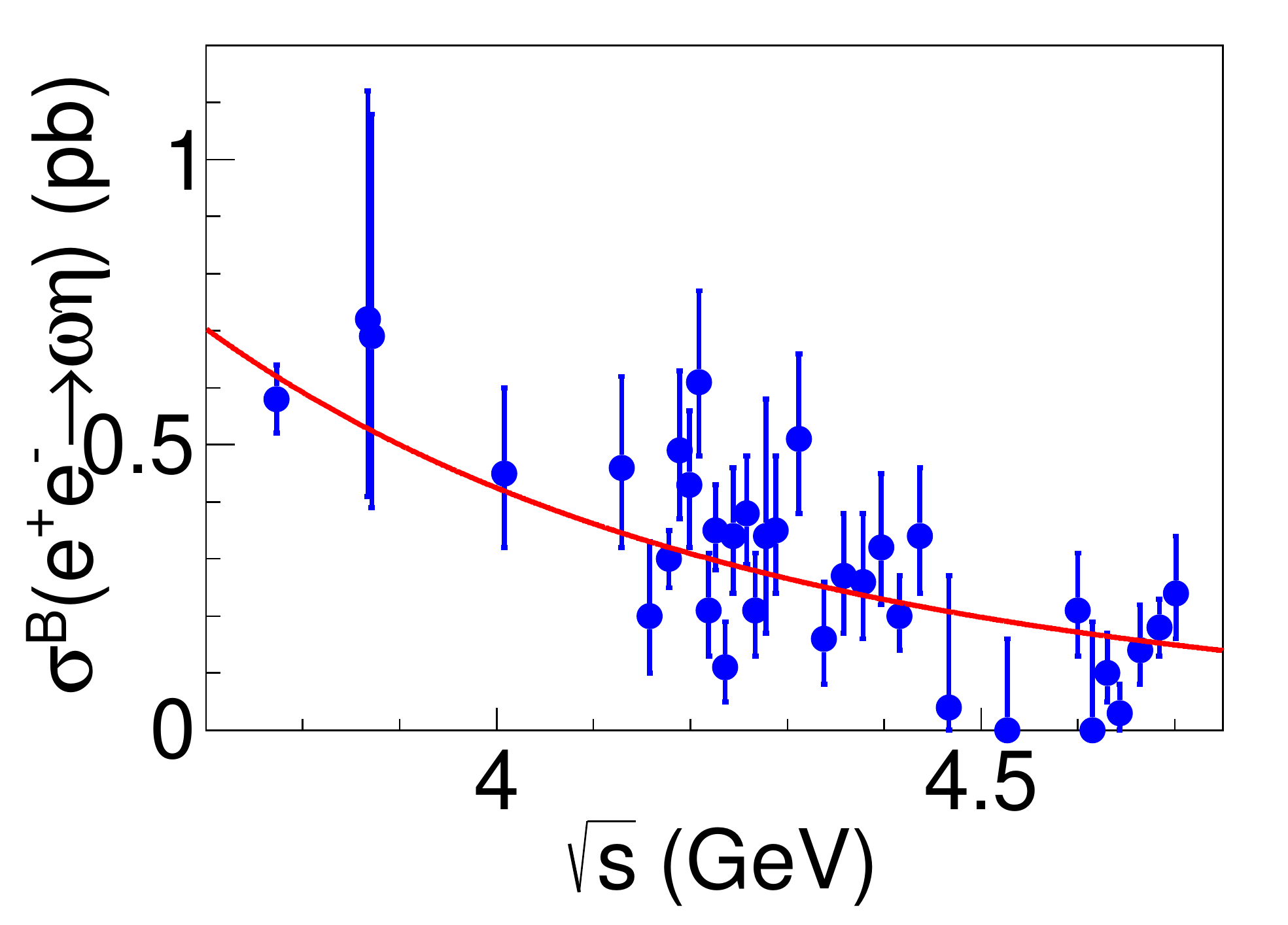}
\end{overpic}
\caption{ Fit to the Born cross sections of $e^{+}e^{-}\rightarrow\omega\eta$ with function $a\cdot s^{-n}$.}
\label{fig:crosseta}
\end{center}
\end{figure}

The Born cross section for $\EE\too\omega\eta$ is calculated using Eq.~\ref{con:sigma1}, where the $N^\text{sig}$ is the
signal yield for $\EE\too\omega\eta$, $\mathcal{B}$ is the product of the branching fractions in the full decay chain $\mathcal{B}=\mathcal{B}(\omega\rightarrow \pi^{+}\pi^{-}\pi^{0})\cdot\mathcal{B}(\pi^{0}\rightarrow\gamma\gamma)\cdot\mathcal{B}(\eta\rightarrow\gamma\gamma)\approx34.78\%$ taken from the PDG~\cite{pdg}. The Born cross section is obtained utilizing the same method described above. Results are listed in Table~\ref{tab:omegapi0etaenergy}.
Figure~\ref{fig:crosseta} shows the energy-dependent Born cross section for $e^{+}e^{-}\rightarrow\omega\eta$, where $n=3.24\pm0.47$, and the goodness of fit $\chi^{2}/n.d.f.=40.45/32$.

The contribution from charmonium(-like) states $\psi(4160)$, $Y(4230)$, $Y(4360)$, $\psi(4415)$ and $Y(4660)$, is investigated similar as the previous part. The corresponding fit function is defined as Eq.~\ref{con:sigma2} in which only $\omega\pi^{0}$ is replaced by $\omega\eta$. The goodness of fit is $\chi^{2}/n.d.f.=40.38/32$ by using a continuum amplitude to fit $\sigma^{\rm{D}}$. Table~\ref{tab:omegaetapi0cross} lists the fit parameters, the statistical significance and $\chi^{2}/n.d.f.$ for the charmonium(-like) states. The systematic uncertainty is not considered in the significance calculation. No significant charmonium(-like) state is found in the process $e^{+}e^{-}\rightarrow\omega\eta$.

\section{Systematic uncertainty}

The systematic uncertainties in the measurement of $\sigma^{\text{Born}}_{\omega\pi^{0}}$ and $\sigma^{\text{Born}}_{\omega\eta}$ originate from the luminosity measurement, the tracking efficiency, the photon detection efficiency, the PID efficiency, the kinematic fit, the $\omega$ mass window, the fitting procedure, the peaking background, the $\theta_{\gamma\gamma}$ requirement, the ISR correction, and the input branching fractions from intermediate states.

The integrated luminosity at each point has been measured with a precision of $1.0\%$ using the Bhabha process~\cite{luminosity}.

The uncertainty from the tracking efficiency is determined to be $1.0\%$ per track~\cite{track1} and the uncertainty in photon detection efficiency is $1.0\%$ per photon~\cite{photon}.

The uncertainty due to PID efficiency is determined to be 1.0\% per pion~\cite{PID}.

The uncertainty due to the kinematic fit requirements is estimated by correcting the helix parameters of charged tracks according to the method described in Ref.~\cite{helix}. The difference between detection efficiencies obtained from MC samples with and without this correction is taken as the uncertainty.

The uncertainty from the $\omega$ mass window is estimated by changing the window range by $\pm$10\%, the maximum difference of final results is taken as systematic uncertainty.

The uncertainty caused by the $M(\gamma\gamma)$ fitting procedure includes the signal shape, background shape and fit range. The original MC-simulated signal shape is replaced by a MC-simulated shape convolved with a Gaussian function, and the difference is taken as the uncertainty from the signal shape. The original linear background function is replaced by a $2^{\rm nd}$-order Chebychev polynomial or a constant one, and the maximum difference is taken as the uncertainty from the background shape.
The fit range is varied by $\pm$10\% at both boundaries, and the maximum difference is taken as the uncertainty from the fit range. The uncertainties from these last three sources are added in quadrature and taken as the total uncertainty from the fitting procedure.

In estimating the systematic uncertainties caused by the $\omega$ mass window and fitting procedure for the channel $e^{+}e^{-}\rightarrow\omega\eta$, the data in all energies are combined together due to the poor statistics.

The uncertainty caused by the peaking background subtraction for the channel $e^{+}e^{-}\rightarrow\omega\pi^{0}$ includes the $\omega$ sideband region and fitting procedure. The $\omega$ sideband regions are shifted by $\pm 6.5$  MeV/c$^2$ (10\% of their widths), and the largest difference of final results is taken as the systematic uncertainty. The uncertainty of the fitting procedure is estimated by the same method as that of $M(\gamma\gamma)$ fitting procedure.

The uncertainty due to the $\theta_{\gamma\gamma}$ requirement for the channel $e^{+}e^{-}\rightarrow\omega\eta$ is estimated by an alternative analysis without the cut criterion of $\theta_{\gamma\gamma}<1$ radian, and the difference of final results is taken as the systematic uncertainty. For the data at energies other than 3.773~GeV, the uncertainties are set as those of 3.773~GeV due to the poor statistics.

The uncertainty from ISR correction factor is estimated by changing the power $n$ in function $\sigma(\sqrt{s})=a\cdot s^{-n}$, which is used to describe the line shapes of $e^{+}e^{-}\rightarrow\omega\pi^{0}, \omega\eta$. The continuum exponent $n$ is shifted by $\pm1\sigma$ relative to the nominal value, and the largest difference of the final result is taken as systematic uncertainty.

The uncertainties caused by the branching fractions from intermediate states are taken from the PDG~\cite{pdg}.

Table~\ref{tab:summererror} summarizes all the systematic uncertainties related to $\sigma^{\text{Born}}_{\omega\pi^{0}}$ and $\sigma^{\text{Born}}_{\omega\eta}$ for each center-of-mass energy. The total systematic uncertainty for each energy point
is calculated as the quadratic sum of the individual uncertainties, assuming them to be uncorrelated.

\begin{table*}[htbp]
  \centering
  \caption{Summary of relative systematic uncertainties (\%) associated with luminosity($\mathcal{L}_{\rm int}$), tracking efficiency (Tracks), photon detection efficiency (Photons), PID efficiency (PID), kinematic fit ($\chi^{2}_{5C}$), $\omega$ mass window ($\omega$), $M(\gamma\gamma)$ fitting procedure(Fit), peaking background (Peaking), $\theta_{\gamma\gamma}$ requirement ($\theta_{\gamma\gamma}$), ISR correction factor (ISR) and branching fraction ($\mathcal{B}$). The first values in brackets are for the process $\EE\rightarrow\omega\pi^{0}$, and the second for the process $\EE\rightarrow\omega\eta$. A dash indicates that the systematic uncertainty is not applicable. The last column (Sum) represents the total systematic uncertainty.}
  \label{tab:summererror}
  \begin{scriptsize}
  \setlength{\tabcolsep}{0.7mm}{
  \begin{tabular}{ccccccccccccc}
  \hline
  \hline
  $\sqrt{s}$ (GeV)  \ & \ $\mathcal{L}_{\rm int}$  \ & \  Tracks  & Photons   & PID   & $\chi^{2}_{5C}$  & $\omega$   & Fit   & Peaking   & $\theta_{\gamma\gamma}$   & ISR   & $\mathcal{B}$  & Sum \\
  \hline
  3.773 & 1.0 & 2.0 & 4.0 & 2.0 & (0.3, 0.3) & (0.3, 1.7) & (0.2, 4.8) & (0.1, -) & (-, 0.5) & (1.5, 1.1) & (0.7, 0.8)  & (5.3, 7.3)\\
  3.867 & 1.0 & 2.0 & 4.0 & 2.0 & (0.3, 0.3) & (0.5, 1.7) & (1.1, 4.8) & (0.1, -) & (-, 0.5) & (1.0, 1.7) & (0.7, 0.8)  & (5.3, 7.4)\\
  3.871 & 1.0 & 2.0 & 4.0 & 2.0 & (0.4, 0.3) & (0.2, 1.7) & (0.5, 4.8) & (0.4, -) & (-, 0.5) & (0.6, 1.5) & (0.7, 0.8)  & (5.1, 7.4)\\
  4.008 & 1.0 & 2.0 & 4.0 & 2.0 & (0.2, 0.1) & (0.4, 1.7) & (1.0, 4.8) & (0.3, -) & (-, 0.5) & (2.8, 0.7) & (0.7, 0.8)  & (5.9, 7.2)\\
  4.129 & 1.0 & 2.0 & 4.0 & 2.0 & (0.2, 0.4) & (0.4, 1.7) & (0.6, 4.8) & (0.2, -) & (-, 0.5) & (0.8, 1.3) & (0.7, 0.8)  & (5.2, 7.3)\\
  4.158 & 1.0 & 2.0 & 4.0 & 2.0 & (0.2, 0.1) & (0.4, 1.7) & (0.3, 4.8) & (0.5, -) & (-, 0.5) & (0.6, 0.8) & (0.7, 0.8)  & (5.1, 7.2)\\
  4.178 & 1.0 & 2.0 & 4.0 & 2.0 & (0.5, 0.3) & (0.1, 1.7) & (1.6, 4.8) & (0.1, -) & (-, 0.5) & (0.6, 0.8) & (0.7, 0.8)  & (5.4, 7.2)\\
  4.189 & 1.0 & 2.0 & 4.0 & 2.0 & (0.3, 0.3) & (0.3, 1.7) & (0.8, 4.8) & (0.2, -) & (-, 0.5) & (0.6, 1.0) & (0.7, 0.8)  & (5.2, 7.3)\\
  4.199 & 1.0 & 2.0 & 4.0 & 2.0 & (0.2, 0.2) & (0.6, 1.7) & (0.3, 4.8) & (0.1, -) & (-, 0.5) & (1.1, 0.9) & (0.7, 0.8)  & (5.2, 7.3)\\
  4.209 & 1.0 & 2.0 & 4.0 & 2.0 & (0.2, 0.3) & (0.6, 1.7) & (0.9, 4.8) & (0.7, -) & (-, 0.5) & (0.7, 0.9) & (0.7, 0.8)  & (5.3, 7.3)\\
  4.219 & 1.0 & 2.0 & 4.0 & 2.0 & (0.3, 0.2) & (0.2, 1.7) & (0.6, 4.8) & (0.8, -) & (-, 0.5) & (1.1, 1.2) & (0.7, 0.8)  & (5.3, 7.3)\\
  4.226 & 1.0 & 2.0 & 4.0 & 2.0 & (0.2, 0.2) & (0.7, 1.7) & (0.6, 4.8) & (0.5, -) & (-, 0.5) & (1.9, 0.5) & (0.7, 0.8)  & (5.5, 7.2)\\
  4.236 & 1.0 & 2.0 & 4.0 & 2.0 & (0.3, 0.2) & (0.3, 1.7) & (1.0, 4.8) & (0.4, -) & (-, 0.5) & (1.0, 0.4) & (0.7, 0.8)  & (5.3, 7.2)\\
  4.244 & 1.0 & 2.0 & 4.0 & 2.0 & (0.3, 0.2) & (0.2, 1.7) & (3.4, 4.8) & (0.6, -) & (-, 0.5) & (0.8, 1.0) & (0.7, 0.8)  & (6.2, 7.3)\\
  4.258 & 1.0 & 2.0 & 4.0 & 2.0 & (0.5, 0.2) & (0.3, 1.7) & (1.0, 4.8) & (0.4, -) & (-, 0.5) & (1.3, 2.2) & (0.7, 0.8)  & (5.4, 7.5)\\
  4.267 & 1.0 & 2.0 & 4.0 & 2.0 & (0.3, 0.2) & (0.7, 1.7) & (0.9, 4.8) & (0.1, -) & (-, 0.5) & (1.2, 1.7) & (0.7, 0.8)  & (5.3, 7.4)\\
  4.278 & 1.0 & 2.0 & 4.0 & 2.0 & (0.3, 0.2) & (1.4, 1.7) & (0.8, 4.8) & (0.1, -) & (-, 0.5) & (0.4, 0.1) & (0.7, 0.8)  & (5.3, 7.2)\\
  4.288 & 1.0 & 2.0 & 4.0 & 2.0 & (0.3, 0.1) & (1.0, 1.7) & (3.0, 4.8) & (0.6, -) & (-, 0.5) & (1.0, 0.8) & (0.7, 0.8)  & (6.1, 7.2)\\
  4.312 & 1.0 & 2.0 & 4.0 & 2.0 & (0.2, 0.3) & (0.8, 1.7) & (0.4, 4.8) & (0.6, -) & (-, 0.5) & (0.9, 0.4) & (0.7, 0.8)  & (5.2, 7.2)\\
  4.338 & 1.0 & 2.0 & 4.0 & 2.0 & (0.3, 0.2) & (0.8, 1.7) & (1.1, 4.8) & (0.6, -) & (-, 0.5) & (0.9, 1.2) & (0.7, 0.8)  & (5.3, 7.3)\\
  4.358 & 1.0 & 2.0 & 4.0 & 2.0 & (0.5, 0.2) & (0.9, 1.7) & (0.5, 4.8) & (0.2, -) & (-, 0.5) & (2.9, 1.0) & (0.7, 0.8)  & (5.9, 7.3)\\
  4.378 & 1.0 & 2.0 & 4.0 & 2.0 & (0.4, 0.2) & (0.6, 1.7) & (0.6, 4.8) & (0.1, -) & (-, 0.5) & (1.2, 1.4) & (0.7, 0.8)  & (5.3, 7.3)\\
  4.397 & 1.0 & 2.0 & 4.0 & 2.0 & (0.2, 0.1) & (0.9, 1.7) & (0.7, 4.8) & (0.8, -) & (-, 0.5) & (1.8, 0.3) & (0.7, 0.8)  & (5.5, 7.2)\\
  4.416 & 1.0 & 2.0 & 4.0 & 2.0 & (0.2, 0.2) & (0.5, 1.7) & (2.3, 4.8) & (0.2, -) & (-, 0.5) & (1.3, 1.0) & (0.7, 0.8)  & (5.7, 7.3)\\
  4.437 & 1.0 & 2.0 & 4.0 & 2.0 & (0.2, 0.2) & (0.9, 1.7) & (0.1, 4.8) & (0.8, -) & (-, 0.5) & (1.3, 0.5) & (0.7, 0.8)  & (5.4, 7.2)\\
  4.467 & 1.0 & 2.0 & 4.0 & 2.0 & (0.3, 0.2) & (1.5, 1.7) & (3.1, 4.8) & (1.0, -) & (-, 0.5) & (0.7, 0.4) & (0.7, 0.8)  & (6.2, 7.2)\\
  4.527 & 1.0 & 2.0 & 4.0 & 2.0 & (0.4, 0.1) & (3.2, 1.7) & (0.1, 4.8) & (0.1, -) & (-, 0.5) & (1.1, 0.7) & (0.7, 0.8)  & (6.1, 7.2)\\
  4.600 & 1.0 & 2.0 & 4.0 & 2.0 & (0.4, 0.3) & (1.3, 1.7) & (1.4, 4.8) & (0.7, -) & (-, 0.5) & (1.4, 1.0) & (0.7, 0.8)  & (5.6, 7.3)\\
  4.615 & 1.0 & 2.0 & 4.0 & 2.0 & (0.3, 0.2) & (1.3, 1.7) & (2.3, 4.8) & (2.0, -) & (-, 0.5) & (1.8, 1.2) & (0.7, 0.8)  & (6.3, 7.3)\\
  4.630 & 1.0 & 2.0 & 4.0 & 2.0 & (0.3, 0.2) & (1.0, 1.7) & (0.3, 4.8) & (0.2, -) & (-, 0.5) & (0.1, 0.5) & (0.7, 0.8)  & (5.2, 7.2)\\
  4.643 & 1.0 & 2.0 & 4.0 & 2.0 & (0.3, 0.3) & (0.8, 1.7) & (1.6, 4.8) & (0.6, -) & (-, 0.5) & (1.9, 1.3) & (0.7, 0.8)  & (5.7, 7.3)\\
  4.664 & 1.0 & 2.0 & 4.0 & 2.0 & (0.3, 0.2) & (1.4, 1.7) & (0.9, 4.8) & (1.1, -) & (-, 0.5) & (0.5, 1.2) & (0.7, 0.8)  & (5.5, 7.3)\\
  4.684 & 1.0 & 2.0 & 4.0 & 2.0 & (0.4, 0.3) & (0.7, 1.7) & (1.1, 4.8) & (0.5, -) & (-, 0.5) & (1.3, 0.8) & (0.7, 0.8)  & (5.4, 7.2)\\
  4.701 & 1.0 & 2.0 & 4.0 & 2.0 & (0.3, 0.2) & (1.0, 1.7) & (1.8, 4.8) & (0.4, -) & (-, 0.5) & (1.8, 1.5) & (0.7, 0.8)  & (5.8, 7.4)\\
  \hline
  \hline
  \end{tabular}}\\
  \end{scriptsize}
\end{table*}

\section{Summary}

We have measured the Born cross sections of two light hadron channels $\EE\too\omega\pi^{0}$ and $\omega\eta$ using data samples collected at BESIII from $\sqrt{s}$ = 3.773 to 4.701 GeV. A power-law function proportional to $s^{-n}$ well describes both line shapes.
No obvious $\psi(4160)$, $Y(4230)$, $Y(4360)$, $\psi(4415)$, or $Y(4660)$ signal is found in the line shapes of $\EE\too\omega\pi^{0}$ and $\omega\eta$.  This indicates a relatively small branching fraction for these resonances into the $\omega\pi^{0}$ and $\omega\eta$ final states.
More exploration of light hadron decay modes will be essential for a further understanding of the charmonium(-like) states.

\acknowledgments

The BESIII collaboration thanks the staff of BEPCII and the IHEP computing center for their strong support. This work is supported in part by National Key R\&D Program of China under Contracts Nos. 2020YFA0406300, 2020YFA0406400; National Natural Science Foundation of China (NSFC) under Contracts Nos. 11875115, 11905179, 11625523, 11635010, 11735014, 11822506, 11835012, 11935015, 11935016, 11935018, 11961141012, 12022510, 12025502, 12035009, 12035013, 12061131003; the Chinese Academy of Sciences (CAS) Large-Scale Scientific Facility Program; Joint Large-Scale Scientific Facility Funds of the NSFC and CAS under Contracts Nos. U1732263, U1832207, U2032110; CAS Key Research Program of Frontier Sciences under Contract No. QYZDJ-SSW-SLH040; 100 Talents Program of CAS; INPAC and Shanghai Key Laboratory for Particle Physics and Cosmology; ERC under Contract No. 758462; European Union Horizon 2020 research and innovation programme under Contract No. Marie Sklodowska-Curie grant agreement No 894790; German Research Foundation DFG under Contracts Nos. 443159800, Collaborative Research Center CRC 1044, GRK 214; Istituto Nazionale di Fisica Nucleare, Italy; Ministry of Development of Turkey under Contract No. DPT2006K-120470; National Science and Technology fund; Olle Engkvist Foundation under Contract No. 200-0605; STFC (United Kingdom); The Knut and Alice Wallenberg Foundation (Sweden) under Contract No. 2016.0157; The Royal Society, UK under Contracts Nos. DH140054, DH160214; The Swedish Research Council; U. S. Department of Energy under Contracts Nos. DE-FG02-05ER41374, DE-SC-0012069.

\clearpage

\section*{The BESIII collaboration}
\addcontentsline{toc}{section}{BESIII Collaboration}
\begin{small}
M.~Ablikim$^{1}$, M.~N.~Achasov$^{10,b}$, P.~Adlarson$^{68}$, S. ~Ahmed$^{14}$, M.~Albrecht$^{4}$, R.~Aliberti$^{28}$, A.~Amoroso$^{67A,67C}$, M.~R.~An$^{32}$, Q.~An$^{50,64}$, X.~H.~Bai$^{58}$
% ...
\\
{\it
$^{1}$ Institute of High Energy Physics, Beijing 100049, People's Republic of China\\
% ...
$^{4}$ Bochum Ruhr-University, D-44780 Bochum, Germany\\
% ...
$^{10}$ G.I. Budker Institute of Nuclear Physics SB RAS (BINP), Novosibirsk 630090, Russia\\
% ...
$^{14}$ Helmholtz Institute Mainz, Staudinger Weg 18, D-55099 Mainz, Germany\\
% ...
$^{28}$ Johannes Gutenberg University of Mainz, Johann-Joachim-Becher-Weg 45, D-55099 Mainz, Germany\\
% ...
$^{32}$ Liaoning Normal University, Dalian 116029, People's Republic of China\\
% ...
$^{50}$ State Key Laboratory of Particle Detection and Electronics, Beijing 100049, Hefei 230026, People's Republic of China\\
% ...
$^{58}$ University of Jinan, Jinan 250022, People's Republic of China\\
% ...
$^{64}$ University of Science and Technology of China, Hefei 230026, People's Republic of China\\
% ...
$^{67}$ (A)University of Turin, I-10125, Turin, Italy; (B)University of Eastern Piedmont, I-15121, Alessandria, Italy; (C)INFN, I-10125, Turin, Italy\\
$^{68}$ Uppsala University, Box 516, SE-75120 Uppsala, Sweden\\
% ...
$^{b}$ Also at the Novosibirsk State University, Novosibirsk, 630090, Russia\\
% ...
}

\end{small}

\end{document}